# The nutshell kings: Why is human space settlement controversial in the first place?

Milan M. Ćirković

*Astronomical Observatory of Belgrade, Volgina 7,*

*11000 Belgrade, Serbia*

**Abstract**. Recent years have seen both a revival of space programs, mostly propelled by private industry's increasing interest, but also the emergence of strong resistance to human space activities on several levels. This is partly a manifestation of a wider counter-Enlightenment *Zeitgeist*, as detectable in other sectors of public life in the West, and partly a reaction against the widespread engagement of the private sector. While it still does not dominate the discourse on space issues, space skepticism is surprisingly wide-ranging and decentralized phenomenon, gathering together such heterogeneous strands of thought from pro-Enlightenment liberals to rabid "deep ecology" activists to philosophical pessimists to antiglobalists of all colors. There has been precious little in way of actively opposing this cultural trend so far, however. While space engineers and entrepreneurs conduct their "business as usual", there are plethora of risks hidden in this cultural climate, esp. if one adopts much repeated (and rarely adequately understood) maxim that "politics is downstream of culture". The present article will review major strands of thought within this "big tent" cultural movement, offer plausible counter-arguments to space skeptics, and outline important cultural and public-outreach work which needs to be done to balance the scales.

> I could be bounded in a nutshell and count myself a king of infinite space, were it not that I have bad dreams.
>
> *Hamlet*, Act 2, Scene 2

## 1. Who's afraid of outer space?

Harvard psychologist Steven Pinker has recently successfully reimagined himself as the main defender of progress and the Enlightenment values in today's largely cynical academia. Amidst fierce polemics with his opponents on both left and right, there's one interesting aspect of (what any reasonable person would consider) progress where he is in complete accord with his opponents: he calls space colonization a "nonstarter", proposed by "naifs".[1] Nothing shouts the extent to which antispace sentiment has become completely normative today louder than the fact that even a strong believer in science and progress like Pinker rejects what certainly could be regarded as the grandest triumph of science and progress—and not a single critic is taking him to task for it.

Imagine somebody coming up to the elders in ancient Greek city states to argue that the then fashionable program of setting up colonies around the Mediterranean is an expression of their naivete. Imagine somebody approaching Prince Henry the Navigator to tell him that all that business with caravels, compasses, cartography, sailing around Africa, etc. is a nonstarter. Imagine—and this is by far the most important analogy—somebody preaching to our early ancestors in the Olduvai Gorge in northern Tanzania that they should not under any circumstances leave their small, known, and cozy (in relative terms) environs and spread around. If we abstract away all difficulties in communication, we can imagine this proto-Pinker to argue, with help of many data graphs and diagrams, that our ancestors would be walking into a disaster if they tried to migrate around the globe, that mortality would be huge, that hardly anyone would survive the migrations, that conditions at most places they would eventually inhabit would be worse than what they already had at the Olduvai, that they will encounter new sicknesses and new predators. Who would have been willing to leave the lush East African homeland to live in places such as Tibet, Alaska, Australian desert, Lapland, Amazonas, etc. Obviously, the very names would not have existed if our ancestors had listened to such voices, Pinker-sophisticated or not. Whether the human species would have survived in such a counterfactual history at the epoch corresponding to our year 2022 is very much open to debate; that its living conditions would have been immensely worse than ours are is eminently *not*.

Thus, we are faced not by one, but by two cultural paradoxes. One is that while recent years have seen strong revival of space programs in many quarters, we also perceive the emergence of strong resistance to human space activities along the "corridors of discourse". Ideological flagships of this negativity surge are the same old utopian visions of extremists on both right and left: sunny future for the chosen nation, race, or class, for which sacrificing the present is justified. Both require a kind of local, parochial, "closed box" view of the world, incompatible with the cosmic future of humanity. During the space race of the 1950s and 60s, such tendencies were present on the margins, but were kept in check by the very ideological polarization of the world. Leftist extremists, for instance, could scoff at NASA, but could hardly afford not to applaud *Sputnik*, Gagarin's flight, *Soyuz*, Tereshkova, and other achievements of the Soviet space program.

---

[1] Pinker (2018), p. 390.

Another paradox, however, is that even intellectuals who are manifestly pro-science and reason (who often self-identify as inheritors of the Enlightenment project) are still enslaved by manifestly narrow, parochial, and at least implicitly counter-Enlightenment dogmas. Although there were some precursors, notably Hannah Arendt's 1963 essay "The Conquest of Space and the Stature of Man", this is a phenomenon of newer vintage. Arguably, it was *enabled* by the ending of the Cold War and the rise of "other concerns" which decoupled the space issues from daily politics and relegated them to the lowest division of human projects.

This is partly a manifestation of a wider counter-Enlightenment *Zeitgeist*, as obvious in other sectors of culture and public life in the West, and partly a reaction against the widespread engagement of the private sector in space. In sharp contrast to some of the previous epochs in which important thinkers developed what could be called the cosmic vision of humanity (e.g., the era of H. G. Wells, Olaf Stapledon, the Russian cosmists, J. B. S. Haldane, J. D. Bernal, etc.), the relevant ideas and memes are now transmitted at shockwave velocity and impact—or, as is here the case, confuse—much larger segments of the public opinion, electorate, and policy-makers.

Arguably, various brands of space skepticism have already reached a rather shrill and hysterical tone in many circles. In the media, it is exemplified by two preposterous instances from *The Atlantic* (which had enthused about the Moon landings back then): one of them being a pompous, ireful, and self-righteous attack targeting *a poster and a joke* (Kriss 2017), the other exhibiting the level of geocentrism coupled with pettiness of concerns rarely seen since the trial of Galileo before the Roman Inquisition in 1633 (Stirone 2021). And the pitch of academia manifests itself on a spectrum from calls for excessive caution[2] to fanciful scare-mongering about future Star Wars, Dooms and Glooms which space settlement will allegedly entail[3] to often hysterical political denunciations of space settlement as an evil imperialist or racist or Big Pharma plot, depending on the favorite set of politically-incorrect scarecrows at any given moment.[4]

While space skepticism has not yet reached the level of prevalence some other cultural movements command, it is arguably expanding its reach – and even crosses-over with other major trends, notably climate anxieties, as testified for instance by a recent blockbuster *Don't Look Up* (2021). Existing data on the lack of public support for space

---

[2] Weinberg (2013); Rees (2018); Szocik (2019). Even notable science-fiction (!) scholars and authors occasionally denounced space settlement in this manner; cf. Westfahl (1997); Stross (2007).
[3] Torres (2019); Kovic (2021).
[4] Slobodian (2015); Vidaurri (2019); Gunderson, Stuart, and Petersen (2021).

activities, sparse as such studies have been so far, are rather solidly on the downside of the issue.[5]

In the rest of this article, I shall offer a plethora of plausible counter-arguments to space skeptics before outlining important cultural and public-outreach work which needs to be done to balance the scales. Among topics which are impossible to cover in this format it is important to mention two: the widely varying *motivations* of individual authors opposed to the human space settlement, which is a topic for (at least) book-length treatment in and of itself, and the consequences of possible existence and properties of extraterrestrial life on our envisioned space settlement prospects. Both are excessively complex topics which require a multidisciplinary analysis which goes far beyond the scope of the present study.

**2. Heavy on emotions, light on logic**

Will space expansion really "diminish the stature of Man [*sic*!]" as Arendt would have us believe? Many modern opponents of space exploration and settlement couldn't resist the temptation of latching onto Arendt's deservedly huge prestige. It is of foremost importance, however, to understand precisely what Arendt did and did not say. The single most important point of her essay, originally published in 1963 and written somewhat earlier, is that the "conquest of space" in the title is *not* to be taken literally; it pertains less to specific space technology/astronautics (there is no indication that she was interested in the topic) than to the general Enlightenment project of melioration through rational, scientific understanding and consequent reductive explanations. It could have equally read "the conquest of matter," "the conquest of geography," or, perhaps most appropriately, "the conquest of biology".

Without entering Arendt's central concern that the effort to reduce all human accomplishment to mere "biological process" will cause our pride to disappear, and will ultimately threaten to lower/destroy the stature of (hu)man, we can see that the role of space is indeed quite peripheral and immaterial here. One is tempted to ask in what sense does, for instance, deep-sea diving lower the stature of (hu)man, since the "conquest of the seas" also offers a technological fix for transcending our natural limitations. As members of species which evolved in a very small part of East Africa, comprising about *6 millionths* of Earth's land real estate, with perhaps a millionth of today's population at some point, to spread all over the globe (including many hostile locales) it is highly ironic to charge "the conquest of space" with adverse consequences. Without the conquest of space, we would not have left the Olduvai Gorge and the Great Rift Valley.

---

[5] Launius (2017). Note the gloomy conclusion of this ambitious study: "A general public lack of support for expending many dollars on spaceflight has been a fundamental reality of NASA since its beginning. It is not changing, and probably not changeable, in the predictive future. Accordingly, NASA's quest for human spaceflight's popular appeal remains an elusive goal." Apart from the emergence of New Space in recent years, there is no clear data on any radical change in this respect since 2017.

Clearly, both "stature" and "dignity" of humans are emotion-laden, fluid, and subjective terms in most of the construals used. One notices that until very recently various authors argued that in-vitro fertilization or organ transplants violate human dignity; many continue to hold similar views with respect to genetically modified foods, stem cell research, or even contraceptives. The Roman Catholic church, for example in Poland, continues to hold IVF "immoral" and "gruesome".[6] While it was remote from Arendt's ideas and concerns, it is worrying to notice that in the course of the last half-century, the insistence on human dignity has become a province of religious zealots and associated radical theologians.

Of course, one should notice that in the circles of bioethics and medical ethics, there is a growing tendency to downplay or avoid the concept of dignity as vague or misleading, and to affirm the ethical positions not making use of it ("undignified ethics"[7]). A great deal of progress, as history of science tells us in no uncertain terms, consists exactly in the rejection of vague and useless concepts in favor of those which are more tangible, quantifiable, and informative.

On the Arendt bandwagon we find, strangely, both Leon Kass-like figures fighting to return religious thinking into mainstream politics, culture, and strategic thinking, as well as those who hold the "deep ecology" view that humankind is a scourge on Earth. Bizarrely enough, these two seemingly antithetical strands of space skepticism – humanist and antihumanist – come together in these confused times (similar to the well-known phenomenon that both left- and rightwing extremists now regularly join forces to protest "global capitalism", free markets, and other liberal cosmopolitan issues). Some examples will be given below.

The tracts of space skeptics abound with such an appeal to emotion. Thus the philosopher Phil Torres (2018, p. 78): "if such individuals were to gain access to a 'doomsday machine' of some sort they would have sadistically, suicidally, and gleefully used it to annihilate their conspecifics" – while the consequences are telling enough and "suicidally" seems reasonable description, one wonders how would the author know about "sadistically" and "gleefully"; if it is not the case, of course, that the emotionally loaded adverbs are used for pure rhetorical effect. In the same category one may find

---

[6] Radkowska-Walkowicz (2018).
[7] Cochrane (2010).

Torres's invoking of an – obviously overblown and already proved to be flawed at the time of writing – scenario by Ray Kurzweil in which "ecophages destroy the entire biosphere of Earth within ~90 minutes" (p. 82). Fear-mongering reaches its unsurpassable apex with the following sentence: "Consider the *billions and billions and billions* of populations that could come to occupy the universe with 10 trillion galaxies and $10^{24}$ stars, each with its own traditions, boasting of weapons that could destroy entire galaxies or even the entire universe" (*ibid.*, emphasis in the original). Indeed, once one reaches these commanding heights of emotional demagoguery, where else to go? We need not be unduly alarmed by "weapons that could destroy entire galaxies or even the entire universe", which are bad science-fiction nonsense if ever there was one.[8] Rather, one is impressed how one Torres in Philadelphia, PA, of 2018 perceives the problems facing inhabitants of $10^{24}$ stars in the AD 1 billion (or so) better than… all those inhabitants themselves. Perhaps it is "their own traditions" which prevent them from seeing wisdom; if they only were smart enough to accept Torres's tradition, everything would have been just fine. I shall return to the underlying Whiggish-like fallacy in the next section.

In contrast to Torres's wide cosmological canvas, Lynda Williams engages in rather short-term fear-mongering when she asserts without evidence that "human missions to Mars do not guarantee the survival of the species, but rather, only the death of any member who attempts the journey" (Williams 2010, p. 6). It is garnered by some quasi-religious appeal to emotion: "If we accept the inevitability of the destruction of Earth and its biosphere, then it is perhaps not too surprising that many people grasp at the last straw and look toward the heavens for solutions and a possible resolution." (*Ibid.*, p. 8) To cite another example in the same vein, the ethicist and entrepreneur Marko Kovic in his comprehensive analysis of the risks of space settlement does not escape fear-mongering in the very abstract of the paper claiming that "it is uncertain whether meaningful space colonization governance can be established in the near future, and before it is too late." Too late for what exactly? Abstract ends there and it is only vague sketched in the paper itself (one option is establishment of "reactionary colonies" which is in itself a biased concept to which I shall return). The very phrase "before it is too late" is almost always a signpost of political demagoguery.[9] The emotional state of false urgency is paradigmatic of the wider concerns of the space skeptics.[10]

---

[8] For elaboration of this embarrassingly trivial point, see Ćirković (2019), sections 3 and 4. See also Torres's popular writings, e.g., https://nautil.us/why-we-should-think-twice-about-colonizing-space-7525/ (last accessed July 3, 2022).

[9] E.g., when used by Donald Trump in a tweet of September 17, 2020, 11:56:59 (https://www.presidency.ucsb.edu/documents/tweets-september-17-2020, last accessed July 3, 2022) or equally repulsive by Bernie Sanders in the foreword and blurb to Teachout (2020).

[10] Note that there is no necessary symmetry in this respect between proponents and opponents of space settlement. If space settlement is regarded as a natural and reasonable evolutionary extension of

A rhetorical strategy similar to the appeal to emotion is assigning guilt by association. When the anthropologist John W. Traphagan discusses the assumptions of a pro-space settlement author, he writes: "A key component of the problem here is that assumptions… are built from the perspectives of wealthy intellectual elites living in wealthy industrial societies – and perhaps not coincidentally living in Western societies that have been historical perpetrators of both colonialism and imperialism" (Traphagan 2019, p. 48). Apart from the usual blaming game, the "not coincidentally" rhetorical device serves to underscore not only politically incorrect provenance, but also to insinuate a *causal link* between pro-space arguments and the perceived "sins of fathers". Which is additionally galling coming, as it does, in a very brief paper affirming tribalism and ethnocentrism in the debates surrounding the human space settlement.

Philosophical pessimism as entailing space skepticism is exemplified by philosopher Robert Klee:[11] because we allegedly have no cosmological future, we should not bother to become a cosmic civilization in the first place. Apart from being a self-fulfilling prophecy peppered by rampant confirmation bias, this makes about as much sense as Woody Allen's famous conclusion (in *Annie Hall*) that doing homework makes no point because cosmology tells us that the universe is expanding. And it is served with pretensions about being guided by *humanist* philosophy – at least Allen's alter ego in the movie was 9 years old.

### 3. Playing fast and loose

While space skeptics are in general prone to label their opponents as adhering more to science-fiction than to science, their own hypocritical use of SF ideas and themes passes mostly under the radar. Thus, when Torres assumes that (post)humans will one day develop "weapons that could destroy entire galaxies,"[12] he is engaging in rather unreasonable SF speculation. There are no indications whatsoever that such superweapons are physically possible. And even if they are possible within some contrived philosophical sense of the "fullness of time", there is no need to worry about it, since there are literally thousands of more probable catastrophic scenarios which could occur naturally and which bear no connection to the issue of space settlement.

---

humanity's ecological niche and the best chance or even precondition for human survival and flourishing—as it is assumed in the present manuscript (and as it is argued by most space settlement advocates)—there is asymmetry in treating emotional appeal to the issue. To receive appeals to emotion equivocally on both sides would be similar to equivocating between appeals to emotion in, say, the debates about slavery, women voting rights, or climate change.

[11] Klee (2017).
[12] Torres (2019), p. 82.

While the notion of such superweapons seems in accordance with the laws of physics (barely), it is implausible that any actor could really build one within the realistic constraint of space, time, and energy available in our real universe. In such cases of extreme improbability, the better bet would perhaps be that our understanding of the laws of physics is actually incomplete and that such astronomically improbable events are, in fact, prohibited. Of course, this does not mean that much smaller and realistic weapons could not be deployed in an interplanetary or interstellar conflict. However, for anyone except a commited nihilist, worrying about such wars and weapons of the distant future would be quite hypocritical, taking into account the plain fact that present-day humans already possess weaponry capable of destroying our planetary habitat. This *straightforward threat* will certainly remain, to our detriment, as long as humanity is limited to one planetary habitat.

A barely coherent panicking such as Torres's over "galaxy-destroying superweapons" is hardly alone in this respect. A similar case is made by ethicist and entrepreneur Marko Kovic, who decries potential "reactionary colonies" humans could possibly establish in space[13] – as if lecturing our descendants centuries or millennia in the future on a present-day ideological basis is anything but unbelievably arrogant presumption. This is a kind of inverted Whiggish fallacy in historiography: while Whig historians castigated people of past epochs for their alleged sins against the present-day political and moral standards, the space skeptics choose to castigate *future* people for their alleged sins against the present-day political and moral standards. If the original Whiggish fallacy is a blind alley of hubris and myopia, the inverted version is doubly ludicrous.

The entire cottage industry of Hobbesian virtue signalling as applied to the distant and likely postbiological future is devoid of sense and rationality. (Note in passing that the entire set of Hobbesian arguments has historically been antithetical to the "human dignity" arguments, Hobbesian materialism being dismissive of the very notion of special "dignity" accorded to human biological machines. This is just another example of internal conflict or incoherency inherent to the patchwork quilt of space skepticism.)

The insistence of Tomasik, Torres, Kovic and some other authors on suffering risks to be allegedly caused by space settlement sounds wise and noble at a glance – until one realizes that it is grounded in a very narrow ethical framework (negative utilitarianism). Since very many other positions are possible and legitimate in moral philosophy, it is clearly dishonest to downplay the fact that adherents of deontology, virtue ethics, contractualism, positive utilitarianism or pragmatic ethics, among others, are not obliged to follow skeptical arguments based on the suffering risks. Especially when the arguments are inherently speculative. Negative utilitarianism is very much a minority view and to insist that it sets the ethical standards for the grandest human endeavors is similar to insistence that, e.g., Mormons or Sikhs set standards of religious observance for all humankind.

---

[13] Kovic (2021), p. 9.

Another popular anti-space canard is that we do need space research, but "robots are better" at it. Proponents fail to address questions such as: *How could robots be better on Moon, Mars or Titan – and not on Earth?* If they are fully capable of such sophisticated activity as space research, how pray tell comes about they are not fully capable of rearing children, composing operas or making political decisions? Or, perhaps, how could not robots be better than humans in writing anti-space tracts? There is no obvious metric on which space research is easier or less complex than those tasks (or myriad others). Note that the effectiveness of this argument is completely orthogonal to the motivation in presenting it: while it is clear that many authors (e.g., Weinberg, Schwartz) are motivated by sincere humanist desire to prevent human deaths in space – which are perhaps as unavoidable as human deaths on high seas ever were – it is easy to argue that many alternative motivations are in play within the wide spectrum of anti-space activism: from speciesism to simple laziness and extreme risk-aversion. As emphasized in the introductory section, the detailed motivational analysis of the anti-space arguments would require a book-length treatment and is beyond the scope of the present manuscript.

In a sense, the robot argument is a *reductio ad absurdum* of space skepticism. Starting from allegedly humanist premises, space skeptics end up affirming the obsolescence of humanity. While there can be no doubt that some particular aspects of space research are optimal for robotization, to argue that this applies to the entire spectrum of space research activities is a dangerous antihuman nonsense. At best, it invokes a *Blade Runner*-like future of speciesist discrimination and police authoritarianism necessary to keep robots "Off World". At worst, it is a prelude to Čapek's *R.U.R.*, where the superior robot-based economy leads to unavoidable extinction of humanity on Earth (and, since they failed to settle other worlds, everywhere).

**4. Pars pro toto**

Most space skeptics argue "only" for *delaying* human space exploration/colonization, as an allegedly "realist" middle ground between proponents of the cosmic future of humanity and those who would burn them at stake as heretics. The argument suggests that *right now* it is not a good time for space settlement, but at some point in the future the time will be just right.[14]

---

[14] Weinberg (2013), Schwartz (2019), or Szocik (2019).

When exactly, though? Since there is no known "natural timescale" for any particular occurrence in cultural evolution, it is hard to assess this argument. How big delay is "sufficient"? Or "optimal"? Or "desirable"? The answer "until we solve all problems on Earth" is nonsense – there will *always* be some problems on Earth. The fact that I am writing and you are reading this now (instead of working directly on space science/technology), could be construed as supportive of *some degree* of delay. At the other extreme, someone who would claim that human space activities should be postponed until the Sun enters the red giant phase, should not be taken seriously in the first place.[15] Usually, we are seeing vague construals such as "it is a task for the distant future" or empty platitudes such as "once we have figured out how to make life on Earth work in… sustainable way".[16]

There is no structural difference between the argument for delaying space colonization and the infamous argument for delaying climate action. In both cases there is a gasping presumption that we shall *inevitably* be better off in the future than we are now; specific climate change mitigation measures which are expensive today will, allegedly, become much cheaper in the future, so we should undertake them 20 or 200 or 2000 years from now, say. Doing otherwise would be tantamount to taking from the poor (present) to give to the rich (future). Irresponsibility and naïveté of such a view defy belief. Adverse consequences of climate change are what already threatens – to a significant degree – global economic prosperity today, which is a precondition for the decrease in price tomorrow. While the question which model of temporal discounting best represents reality still lacks a clear answer, nobody seriously doubts the necessity of such discounting in the futures studies (e.g., Tonn and Stiefel 2014).

Why should it differ for human space settlement? Insofar as the latter is regarded as decreasing the risk of human extinction and increasing the robustness of human civilization, the parallel holds. While it is reasonable to expect that – *barring crises and catastrophes*! – the direct cost of human space settlements will decrease in the future, one is still entitled to the same scepticism toward it as in the case of climate change: leaving the brunt of work to one's descendants betrays cowardice in face of responsibility. Coupled with corrosive influence of social constructivism on the public perception of relevant science (Hansson 2020), such attitude abounds in global dangers and threats.

The delay argument is usually bolstered by a *pars pro toto* fallacy: skeptical discourses which are focused on a single small facet of the entire vast front of space expansion. In cartoonish terms this reasoning can be represented as follows:

---

[15] Thus Williams (2010, p. 4): "So yes, we are doomed, but we have five billion years, plus or minus a few hundred million, to plan our extraterrestrial escape." Nice to note that there are central-planners who openly entertain the Five Billion Year Plan.
[16] Weinberg (2013); Williams (2010).

**(A)** Proponents of space settlement often express their desire/goal/intention to settle Mars.

**(B)** Long-term habitation of Mars is impossible/unfeasible/undesirable.

Hence,

**(C)** Proponents of space settlement have no case.

This is not only formally wrong in the sense that the premises do not entail the conclusion, but willfully misleading on several counts. Obviously, Mars is a very, very, very small part of the surrounding universe. For all worthiness of the pursuit of settling Mars, even if the premise **(B)** is granted, this certainly does not exhaust the list of possible targets for space settlement, especially so many decades after the epochal work of Gerard O'Neill, who persuasively argued for entirely artificial habitats, or those constructed by adapting small bodies (asteroids/cometary nuclei) for human habitation. Therefore, even if Mars or the Moon or Ganymede turn out to be impractical for settlement, many options remain, in the main Asteroid Belt, in the Kuiper belt, among NEOs, and in convenient orbits around Earth or other bodies. Often shrill criticism of terraforming belongs here as well: whatever intrinsic merits or demerits of terraforming there are, its practice is and will likely forever be a very, very small part of the space settlement advocacy.

This side of the story has nothing to do with the intrinsic likelihood of the premise **(B)**. I find that likelihood low, and we may have various opinions on that, but the onus of proof is on skeptics. Again, imagine a prophetic voice telling one of our ancestors, fresh out of the Olduvai Gorge, that they should not migrate, since it is so cold in (what will become) Scandinavia or Alaska, so wet in Japan or Brazil, so dry in Sahara or the Australian interior, etc.

**5. Conclusions: the poverty of skepticism**

In the course of this paper, I have surveyed a wide and often wild array of arguments of space skeptics; as far as possible I have refrained from presenting the positive case for space, since many authors have done that in more masterly and comprehensive manner (e.g., O'Neill 1977; Tsiolkovsky 2004; Beech 2007; Musk 2017, 2018; Green 2019; Sivolella 2019; Zubrin 1999, 2019). At the present level of knowledge and understanding, almost all skeptical arguments are either too vague/emotional to be properly engaged, or fail to meet serious critical scrutiny, or are not, in fact, opposed to the space expansion generally, criticizing only a particular objectionable segment of it. In addition, space skepticism lacks internal coherency; we are dealing with a hodge-podge of various complaints and peeves instead, often grounded in totally opposed and incompatible ideologies ("humans are planetary cancer which should not be allowed to spread" vs. "humans are too precious to be risked or exploited in space settlements"). There are, however, still two key questions to address.

**Do we need to worry?** One may think that all the media/academic hot air does not really impact the progress of space technology and development. While NASA and the small bunch of its analogs may hire marketing teams to counter the skeptic worries (often in a rather hypocritical manner, hence confirming—instead of dispelling— paranoiac and conspiratorial assertions of the critics), their piece of the cosmic cake is getting smaller. Most engineers and mission personnel may endorse the "business as usual" attitude. This would be a serious mistake, similar to the one committed by most in the natural sciences when faced with the onslaught of postmodernism, identity politics and critical theory nonsense in academia: it took only two decades from the Sokal Hoax to reach the stage in which NASA organizes *Mission: Equity* and gets involved in re-christening transneptunian objects in agreement with the newest pol-correct fads (meanwhile debating merits of the scandalous $4.1 bn price tag for *a single Artemis launch*[17]), while Cornell University offers shockingly foolish nonsense in "racial cosmology" as a course.[18] So, recent experience demonstrates that stupid *is* contagious—and is moving in the direction of science and engineering. The fact that space skepticism is not sufficiently dominant in the public and cultural sphere to directly obstruct specific programs and activities should not lull us into complacency about its destructive and confusing potential.

One may ask a pertinent question about the relevance of the public sphere and cultural climate in the epoch in which most space industry passed from the government to private actors. Clearly, it is unrealistic to assume that governments will relinquish their role in directing and regulating our space efforts any time soon. Governments are – and are likely to remain for quite some time – significant developers of technologies that form the foundation of new space industry segments and industries. In such situation the main issue is what regulatory pressures and other statist issues are optimal from the point of view of efficient space exploration and settlement. While the space skepticism could arguably be less impactful in the mostly privatized New Space context, it can still poison the public discourse and lead to completely unnecessary and irrational tensions in culture and society. So, yes, we need to worry.

---

[17] https://oig.nasa.gov/docs/IG-22-003.pdf, last accessed July 3, 2022.
[18] https://complit.cornell.edu/news/featured-course-black-holes-race-and-cosmos, last accessed July 6, 2022.

**What is to be done?** This is the real crux of the issue. There has been precious little in way of activelly opposing the cultural trend so far. While space engineers and enterpreneurs conduct their "business as usual", there are a plethora of risks hidden in this cultural climate, esp. if one adopts the much repeated (and rarely adequately understood) maxim that "politics is downstream of culture". We need more free debate on these issues – which is becoming more and more difficult in the current authoritarian atmosphere. On the tactical level, there *are* things space advocates should learn from the skeptics: appeals to emotions, social slant, humanist rhetoric with the advantage that space advocacy is, for a change, a *truly humanist* program. Education in STEM is of paramount importance – and there is simply no excuse for policy discussions on any level by societal actors often without any exposure to STEM fields. More clear critical thinking will contribute to suppression of the fashionable "cosmic pessimism" – often feeding on the ignorance – which is at best confusing and at worst extremely dangerous as far as the future of humanity is concerned.

Finally, a change in the overall mindset is highly desirable. It is only *our own* intellectual inertia and laziness which makes us short-sighted and devoid of true vision. In a parable due to Kierkegaard and retold by Borges, Danish priests initially offered indulgences for undertaking North Pole expeditions. Since such expeditions are difficult and require considerable courage and effort, priests gradually relaxed the conditions under which they defined participation in such an expedition. At the end, they concluded that *any travel* whatsoever, like sailing from Copenhagen to London, or traveling to a Sunday picnic, qualifies as a true North Pole expedition. This is a perfect metaphor for the global state of self-indulgence, complacency, and cowardice (with honorable exceptions) about the cosmic future of humanity. While a Tsiolkovsky, a Stapledon, or an O'Neill boldly proclaimed our cosmic destiny a century ago, today's NASA and related organizations engage in postmodern virtual nonsense. The ongoing pandemic of SARS-CoV-2 should, if anything, teach us about fragility of the terrestrial civilization under even mild threats of either natural or anthropogenic origin.

As Shakespeare understood, our nutshell kings would be perfectly fine – except for the bad dreams they're having. Their bad dreams, full of jealousy (and pathological politics of jealousy), geocentrism, anthropocentrism, parochialism both cosmic and terrestrial, short-termism morphing into reckless short-sightedness, politically correct virtue signaling, fake pragmatism and false precaution, as well as myriad other tools of *concern trolling*, are certainly a thing in today's world. Freud (and Otto Rank and many subsequent psychoanalysts/existentialists) would certainly connect such bad dreams with the *trauma of birth*; going a single step further than Tsiolkovsky, one may evoke *Earth not as the cradle, but as the womb* of an emerging cosmic civilization. Leaving it is traumatic and violent. No wonder the nutshell kings are horrified by the prospect. They would prefer not to be born.

Those nightmares should be contrasted with noble dreams of the great visionaries of the world, of Fedorovs and Stapledons, of Tsiolkovskys and Buckminster Fullers, of Teslas and Dysons, of Clarkes and Kubricks, of Lems and O'Neills, of Sagans and Kurzweils and Sandbergs and Musks – and the unknown leaders of migrations from Olduvai. Great minds not just of Earth, but of the universe itself, the true cosmopolitans (= citizens of the cosmos), inheritors of truly inspiring evolutionary traditions of expanding into available ecological and creative niches. This wonderful planet is indeed too minuscule to confine those noble dreams which indeed belong to the entire universe.

Every one of us is free to choose the kind of dreams to adhere to.

The nutshell will be cracked.

**Acknowledgements**. Two anonymous referees have kindly supplied useful comments resulting in significant improvement upon an early version of this manuscript. Pleasant discussions with Milica Banović, Keith Mansfield, Slobodan Popović, Srdja Janković, Mark Walker, Paul Gilster, Bojan Stojanović, Slobodan Perović, and Claire Berlinski are also hereby acknowledged.